\documentclass[fleqn,twoside]{article}
\usepackage{espcrc2}
\usepackage{amsmath}

\usepackage{graphicx}
\usepackage[figuresright]{rotating}

\title{Neutrino magnetic moment: a window to new physics}

\author{A. Studenikin\address{Department of Theoretical Physics,
        Moscow State University, 119991 Moscow, Russia}
        \thanks{e-mail:
        studenik@srd.sinp.msu.ru}}

\begin{document}

\begin{abstract}
    A short review on a neutrino magnetic moment is presented.
\end{abstract}

\maketitle

{\it Introduction.} Experimental and theoretical studies of flavour
conversion in solar, atmospheric, reactor and accelerator neutrino
fluxes give strong evidence of non-zero neutrino mass.  A massive
neutrino can have non-trivial electromagnetic properties
\cite{MarSanPLB77LeeShrPRD77FujShrPRL80} . For a recent review on
neutrino electromagnetic properties see \cite{GiuStuPAN09}.

The neutrino dipole magnetic moment (along with the electric dipole
moment) is the most well studied among neutrino electromagnetic
properties.  The effective Lagrangian, that is in charge of a
neutrino coupling to the electromagnetic field, can be written in the
form
\begin{equation}\label{Lagr_sigma_F}
L_{int}=\frac{1}{2}{\bar \psi}_i \sigma_{\alpha \beta}(\mu
_{ij}+\epsilon_{ij} \gamma_{5}) \psi_j F^{\alpha \beta}+ h.c.
\end{equation}
where the magnetic moments $\mu_{i j}$, in the presence of mixing
between different neutrino states, are associated with the neutrino
mass eigenstates $\nu_i$.  The interplay between magnetic moment and
neutrino mixing effects is important. Note that electric (transition)
moments $\epsilon_{ij}$ do also contribute to the coupling.

A Dirac neutrino may have non-zero diagonal electric moments in
models where $CP$ invariance is violated. For a Majorana neutrino the
diagonal magnetic and electric moments are zero. Therefore, neutrino
magnetic moments
 can be used to
distinguish Dirac and Majorana neutrinos (see
\cite{ShrNP82SchValPRD81PalWolPRD82KayPRD82KayPRD84NiePRD82} and also
\cite{GiuStuPAN09} for a detailed discussion).

{\it Neutrino magnetic moment in a minimal extension of Standard
Model.} The explicit evaluation of the one-loop contributions to the
Dirac neutrino magnetic moment  in the leading approximation over
small parameters $b_i=\frac{m_{i}^{2}}{M_{W}^{2}}$ ($m_i$ are the
neutrino masses, $i=1,2,3$), that however exactly accounts for $a_l=
\frac{m_{l}^{2}}{M_{W}^{2}}$ ($l=e,\ \mu , \ \tau$), leads the
following result \cite{PalWolPRD82},
\begin{equation}\label{m_e_mom_i_j}
  \mu^{D}_{ij}=\frac{e G_F m_{i}}{8\sqrt {2} \pi ^2}
  \Big(1 + \frac{m_j}{m_i}\Big)\sum_{l=\ e, \ \mu, \
  \tau}f(a_l)U_{lj}U^{\ast}_{li},
\end{equation}
\begin{equation}\nonumber
f(a_l)=\frac{3}{4}\Big[1+\frac{1}{1-a_l}-\frac{2 a_l}
{(1-a_l)^2}-\frac{2a_l^2\ln a_l}{(1-a_l)^3} \Big],
\end{equation}
where $U_{li}$ is the neutrino mixing matrix. The correspondent
result in the absence of mixing was confirmed in
\cite{Car_RosBerVid_ZepEPJC00,DvoStuPRD04_DvoStuJETP04}. A Majorana
neutrino may also have transition moment of the value
$\mu^{M}_{ij}=2\mu^{D}_{ij}$ (see \cite{GiuStuPAN09} for a detailed
discussion  and references).

For the diagonal magnetic moment of the Dirac neutrino, from
(\ref{m_e_mom_i_j}) in the limit $a_l\ll 1$ the result
\cite{MarSanPLB77LeeShrPRD77FujShrPRL80} can be obtained
\begin{equation}\label{nu_mu_D_ii}
    \mu^{D}_{ii}=\frac{3e G_F m_{i}}{8\sqrt {2} \pi ^2}
  \Big(1 - \frac{1}{2} \sum_{l=\ e, \mu, \tau}a_l\mid
  U_{li}\mid^{2}\Big).
\end{equation}

The magnetic moment for hypothetical heavy neutrino was studied in
\cite{DvoStuPRD04_DvoStuJETP04}.  In particular, it was obtained
\begin{equation}
    \mu_{\nu}={\frac{eG_{F}m_{\nu}}{8\sqrt{2}\pi^{2}}}
\Bigg  \{\begin{array}{c}
    3+{\frac{5}{6}}b, \ m_\ell\ll m_\nu\ll M_W, \\
    1,\ \ \ \ \ \ \  \  m_\ell\ll M_W\ll m_\nu.
  \end{array}
      \end{equation}
Note that the $LEP$ data set a limit on number of light neutrinos
coupled to $Z$ boson.

The numerical value of the Dirac neutrino magnetic moment within a
minimal extension of the Standard Model, as it follows from
(\ref{nu_mu_D_ii}), is
\begin{equation}\label{mu_3_10_19}
    \mu^{D}_{ii}\approx 3.2\times 10^{-19}
  \Big(\frac{m_i}{1 \ eV}\Big) \mu_{B},
\end{equation}
This is several orders of magnitude smaller than the present
experimental limits if to account for the existed constraints on
neutrino masses.

{\it Neutrino magnetic moment in other extensions of Standard Model.}
Much larger values for a neutrino magnetic moments can be obtained in
different other extensions of the Standard Model (see
\cite{KimPRD76BegMarRudPRD78}, the first paper of
\cite{MarSanPLB77LeeShrPRD77FujShrPRL80} and, for instance,
\cite{CzaGluZraPRD99_KukKik0306025}). However, there is a general
problem \cite{9,10,11} for a theoretical model how to get a large
magnetic moment for a neutrino and simultaneously to avoid
unacceptable large contribution to the neutrino mass. If a
contribution to the neutrino magnetic moment of an order $\mu_{\nu}
\sim \frac{eG}{\Lambda}$ is generated by physics beyond a minimal
extension of the Standard Model at an energy scale characterized by
$\Lambda$, then the correspondent contribution to the neutrino mass
is
\begin{equation}\label{mu_Lambda}
\delta m_{\nu} \sim \frac{\Lambda ^2}{2m_e}\frac{\mu_{\nu}}{\mu_B}=
\frac{\mu_{\nu}}{10^{-18}\mu_B}\Big(\frac{\Lambda}{1 \ Tev}\Big)^2\
eV.
\end{equation}
Therefore, a particular fine tuning is needed to get a large value
for the neutrino magnetic moment while keeping the neutrino mass
within experimental bounds.

Different possibilities to have a large magnetic moment for a
neutrino were considered  in the literature (see, for instance,
\cite{VolSJNP88_BarFreZeePRL90}).

{\it Bounds on neutrino magnetic moment.} The constraints on the
neutrino magnetic moment in the direct laboratory experiments so far
obtained from unobservant distortions in the recoil electron energy
spectra. The best upper bounds on the neutrino magnetic moment are
obtained in the recently carried reactor experiments: $\mu_\nu \leq
9.0 \times 10^{-11}\mu_{B}$ ($MUNU$ collaboration
\cite{Dar_eaPLB05}),  $\mu_\nu \leq 7.4 \times 10^{-11}\mu_{B}$
($TEXONO$ collaboration \cite{Wong_PRD07_75_012001}, and $\mu_\nu
\leq 5.8 \times 10^{-11}\mu_{B}$ ($GEMMA$ collaboration
\cite{BedStar_eaPAN07}\footnote{Further improvement of the bound is
expected soon \cite{BedSta13LomCon}.}). Stringent limits  also
obtained  in the solar neutrino scattering experiments: $\mu_\nu \leq
1.1 \times 10^{-10}\mu_{B}$ ($Super-Kamiokande$ collaboration
\cite{Liu_ea_PRL04}) and $\mu_\nu \leq 5.4 \times 10^{-11}\mu_{B}$
($Borexino$ collaboration \cite{BOREXINO_08}).

Note that the global fit \cite{Grim_ea_NPB03_Tor04} of the magnetic
moment data from the reactor and solar neutrino experiments for the
Majorana neutrinos produces limits on the neutrino transition moments
$\mu_{23},\ \mu _{31}, \ \mu _{12} < 1.8 \times 10^{-10}$. Upper
limits on magnetic moments for the muon and $\tau$-neutrino neutrinos
($\mu_{\nu _\mu} \leq 1.5 \times 10^{-10}\mu_{B}$ and $\mu_{\nu
_\tau} \leq 1.9 \times 10^{-10}\mu_{B}$, respectively) were found
\cite{MonPicPul08012643} in an independent analysis
 of the first release of the $Borexino$ experiment data.

It should be mentioned \cite{9} that what is measured in scattering
experiments is an effective magnetic moment $\mu^{exp}_{e}$, that
depends on the flavour composition of the neutrino beam at the
detector located at a distance $L$ from the source, and which value
is a rather complicated function of the magnetic (transition) moments
$\mu_{i j}$:
\begin{equation}\label{mu_exp}\nonumber
\mu_{exp}^{2} = \mu_{\nu}^{2}(\nu_{l},L,E_{\nu})=\sum_{j}\Big|
\sum_{i} U_{li} e^{-iE_{i}L}\mu_{ji} \Big| ^{2}.
\end{equation}
The dipole electric (transition) moments, if these quantities not
vanish,  can also contribute to $\mu^{exp}_{e}$.

A general and model-independent upper bound on the Dirac neutrino
magnetic moment, that can be generated by an effective theory beyond
the standard model, have been derived \cite{11}: $\mu_{\nu}\leq
10^{-14}$ (the limit in the Majorana case is much weaker).

{\it Neutrino magnetic moment  interaction effects.} If a neutrino
has non-trivial electromagnetic properties, notably non-vanishing
magnetic (and also electric (transition) dipole moments or non-zero
millicharge and charge radius), then a direct neutrino couplings to
photos becomes possible and several important for applications
processes exist \cite{RafPRL90}. A set of typical and most important
neutrino electromagnetic processes involving the direct neutrino
couplings with photons is: 1) a neutrino radiative decay
$\nu_{1}\rightarrow \nu_{2} +\gamma$, neutrino Cherenkov radiation in
external environment (plasma and/or electromagnetic fields), spin
light of neutrino, $SL\nu$ , in the presence of medium \cite{SLnu}; \
2) photon (plasmon) decay to a neutrino-antineutrino pair in plasma
$\gamma \rightarrow \nu {\bar \nu }$, \ 3) neutrino scattering off
electrons (or nuclei), 4) neutrino spin (spin-flavor) precession in
magnetic field. Note that resonant neutrino spin-flavour oscillations
in matter were considered in \cite{LimMarPRD88_AkhPLB88}.

The tightest astrophysical bound on a neutrino magnetic moment is
provided by observed properties of globular cluster stars. For a
large enough neutrino magnetic moment the plasmon decay rate can be
enhanced so that a reasonable delay of helium ignition would appear.
From lack observation evidence of anomalous stellar cooling due to
the plasmon decay the following limit has been found \cite{RafPRL90}
\begin{equation}
\Big( \sum _{i,j}\mid \mu_{ij}\mid ^2\Big) ^{1/2}\leq 3 \times
10^{-12} \mu _B.
\end{equation}
This is the most stringent astrophysical constraint on a neutrino
magnetic moment, applicable to both Dirac and Majorana neutrinos.

{\it Conclusion.} There is a huge gap of many orders of magnitude
that does exist between the present limits $\propto 10^{-(11\div
14)}\mu_{B}$ on a neutrino magnetic moment $\mu_{\nu}$ and the
prediction of a minimal extension of the Standard Model. Therefore,
if any direct experimental confirmation of non-zero neutrino magnetic
moment were obtained within a reasonable time in the future, it would
open a window to new physics.

{\it Acknowledgements.} The author is thankful to Gianluigi Fogli for
the invitation to attend the Neutrino Oscillation Workshop and to all
of the organizers for their kind hospitality in Conca Specchiulla.
The author is also thankful to Carlo Giunti for discussions on
neutrino electromagnetic properties.

\end{document}